\documentstyle[12pt,epsf]{article}
\newcommand{\tresj}[6]{\left( \begin{array}{ccc}
                              #1 & #2 & #3 \\
                              #4 & #5 & #6 
                             \end{array}
                      \right)}

\begin{document}
\begin{titlepage}
\thispagestyle{empty}
\begin{flushright}
UG-DFM-3/97 \\
nucl-th/9704022
\end{flushright}  
  \vspace*{5mm}

\vspace*{2cm}

\begin{center}
{\Large \bf Radiative pion capture in nuclei:\\ a continuum
shell--model 
approach}

\vspace{1.5cm}
{\large J.E. Amaro, A.M. Lallena and J. Nieves}

\vspace{.3cm}
{Departamento de F\'{\i}sica
Moderna, Universidad de Granada, \\
E-18071 Granada, Spain}

\end{center}

\vspace{2cm}
\begin{abstract}
The radiative pion capture process in nuclei is approached by
using a continuum shell--model description of the nucleus, 
together with a phenomenological treatment of the two
particle--two hole effects. 
It is found that these effects play an important role to
reproduce the observed experimental photon energy distribution.
This distribution as well as the integrated one
depends significantly on the details of the mean field
potential. This makes this process interesting to investigate 
the nuclear structure dynamics.

\vspace{1cm}

\end{abstract}

\end{titlepage}

\newpage

\setcounter{page}{1}

\section{Introduction}

The radiative pion capture (RPC) in pionic atoms has 
been extensively studied in the past~\cite{Ver75}-\cite{Kri88}.
In this process, the $\pi^-$ sited in a given $nl$ orbit of a 
pionic atom is captured by the nucleus emitting a photon. The 
interest of this reaction is threefold. First, it provides direct
information about the $\pi$-nucleus interaction, responsible for the
formation of the pionic bound state. Second, it proves the nuclear structure,
specifically the details of the proton-distribution, and the nature of
the two particle--two hole ($2p2h$) effects in the nuclear medium. 
Finally, it gives
valuable information on the $\pi^- p \to \gamma n $ reaction on the nucleon 
at very low pion kinetic energies and in this sense it is
complementary to the pion photoproduction reactions. In this paper we
will concentrate on the second aspect mentioned above.

Technically, the most difficult aspect in the theoretical side is 
the evaluation of the sum over final nuclear states. This has made
necessary the use of different approaches such us the closure 
approximation~\cite{Roi85} or the RPA sum rules~\cite{Kri88}. In
the first case, the results show a considerable dependence on the
average photon energy used, a problem which is minimized by the RPA 
sum rules method. However, this two approximations only give 
integrated quantities (e.g. the radiative capture widths and  
branching ratios), but observables such as the photon energy
distribution are out of their possibilities.

More recently, Chiang {\it et al.}~\cite{Chi90} have proposed a 
many--body approach in which both the photon energy distribution and
the radiative capture width can be calculated. Starting from techniques
developed for the $(\gamma,\pi)$ reaction~\cite{Car89}, this method 
permits the use of accurate pionic wave functions and the incorporation
of the medium polarization effects. Besides, the calculations are 
considerably simplified because they are done in infinite nuclear matter 
and applied to finite nuclei by means of the local density approximation
(LDA). Though the obtained branching ratios and widths are in overall 
agreement with the experimental results, the photon energy
distribution does not show
the marked tail of the data at low photon energies. This last problem
is partially due to the limited momentum components available to
the nuclear wave functions obtained from the Fermi gas (FG) model.

Shell--model (SM) calculations do not present this 
deficiency. This was observed
in the pioneering work of Vergados~\cite{Ver75} and Eramzhyam 
{\it et al.}~\cite{Era77} and in the results obtained
by Ohtsuka and Ohtsubo~\cite{Oht78} 
and W\"unsch~\cite{Wun80} within the continuum SM approach. Much 
of this work has been devoted to the analysis of the role played by the
resonances in the reaction and the main conclusion is
that the RPC can be described assuming a dominant resonance 
character for the process.

In this work we want to investigate the dynamics of the RPC reaction 
in more detail by analyzing it
 with a continuum SM recently developed to 
study the quasi--free inclusive electron scattering by nuclei
~\cite{Ama93,Ama94}.
In particular we want to address different questions related to the 
importance of the parameters of the SM potential, to the problem of 
the orthogonality of the single--particle wave functions and to the
effect of the inclusion of the 
particle--hole ($ph$) self--energy in the medium on the  $ph$
propagator. This
latter effect  accounts for the final state interactions (FSI) of
the emitted neutron with the residual nucleus.

The organization of the paper is as follows. In sect.~2 we give the details
of the model used to perform the calculations with the explicit
expressions of  the necessary matrix elements 
placed in the appendix.  In sect.~3
we show the results and comment the main features. In sect.~4 we
summarize our work and draw our conclusions. 


\section{Details of the model}


\subsection{The RPC rate}

In this work we study the following reaction:

\begin{equation}
\left(^A_Z{\rm X}_N \, + \pi ^-\right)_{\rm bound} \, \rightarrow \, 
\gamma \, + \,n\, + \, _{Z-1}^{A-1}{\rm Y}_N 
\end{equation}
where the pion  is captured by the nucleus emitting a
neutron and a photon, being the latter one the only particle detected.

The unpolarized differential photon distribution  
for the capture of a pion, bound
in the orbit $nl$, from a spin zero nucleus is given by~\cite{deF66}:
 
\begin{equation} \label{gamma}
\displaystyle \frac{d\Gamma^{(\gamma)}_{nl}}{dq}(q)
\, = \, \frac{8\pi\alpha q}{2J_i+1}
       \sum_{fJ}\delta(q+E_f-E_i)
       \left[ |\langle f \|T^{\rm el}_J(q)\| i\rangle|^2 \, + \,
                     |\langle f \|T^{\rm mag}_J(q)\|i\rangle|^2 \right] .
\end{equation}
Here it has been taken into account the fact that the nuclear states as well
as that of the pion have good angular momenta, what makes useful to perform
a multipole expansion of the operators involved. In Eq.~(\ref{gamma}), 
$\alpha$ is the fine structure constant, $q$ is the energy of the 
emitted photon, $|i\rangle$ and $| f\rangle$   represent 
the initial (taking into account the pion degrees of freedom) 
 and  final hadronic states of the system and $E_{i,f}$ are their
respective  energies.
Besides, $T^{\rm el}_{JM}$ and
$T^{\rm mag}_{JM}$ are the electric and magnetic multipole operators which are
given by:
\begin{eqnarray} \label{electric}
T_{JM}^{\rm el}(q)&=& \frac{1}{q} \, 
  \int {\rm d}^3 r \, \left\{ \nabla \times \left [
  j_J(qr)\,{\bf Y}_{JJ}^M (\hat{\bf r}) \right] \right\} 
\cdot {\bf J}({\bf r}) , \\
\label{magnetic}
T_{JM}^{\rm mag}(q)&=&
  \int {\rm d}^3 r \,
  j_J (qr)\,{\bf Y}_{JJ}^M (\hat{\bf r})\cdot {\bf J} ({\bf r}),
\end{eqnarray}
where  $j_J$ is a spherical Bessel
function and {\bf Y}$_{JL}^M$ a vector spherical harmonic. Finally,
{\bf J}({\bf r}) is the electromagnetic current operator.

The amplitude for the elementary process $\pi^- p \to \gamma n $,
which determines the hadronic current, depends~\cite{Eri88} 
on the spin of the nucleon 
involved in the process, the photon polarization and the pion and photon 
momenta. However, the average momentum of the bound pion in light and
medium nuclei is small and in this situation it is plausible to
take into account only the so--called Kroll--Ruderman term (which is the
one surviving in the limit of zero pion momentum). Then we have for the
electromagnetic current operator
\begin{equation} \label{current}
\displaystyle
{\bf J}({\bf r}) \, = \, i \sqrt{2} \, \frac{f}{m_\pi} \, 
\left( 1 + \frac{m_\pi}{2M} \right) \,
\sum_{i=1}^A \, t_-(i) \, \phi_-({\bf r}) \,
\mbox{\boldmath $\sigma$}(i)  \delta({\bf r}-{\bf r}_i) .
\end{equation}
Here, the index $i$ runs over the $A$ nucleons of the system
with coordinates ${\bf r}_i$, $f$ is the pion--nucleon 
coupling constant ($f^2/4\pi=0.08$), 
$m_\pi$ is the pion mass, $M$ is the nucleon mass and $\mbox{\boldmath
$\sigma$}$ are spin Pauli matrices. Besides, the isospin
operator $t_-$ transforms a 
proton into a neutron. Finally, $\phi_-({\bf r})$ is the pion field operator
which can be developed in a complete basis as follows:
\begin{equation}
\phi_-({\bf r}) \, = \, \sum_k \, \displaystyle 
\frac{1}{\sqrt{2E^\pi_k}} \,
\phi_k({\bf r}) \, c_k ,
\end{equation}
where ${k}$ runs over all the pionic atom orbits, $\phi_k$ 
is the corresponding pion wave function, $c_k$ is the pion
annihilation operator, 
 and $E^\pi_k$ is the pion energy including its mass.

After substituting the current~(\ref{current}) in Eqs.~(\ref{electric}) and 
(\ref{magnetic}) it is possible to obtain the tensor form of these 
electromagnetic operators and  calculate the matrix elements involved
in Eq.~(\ref{gamma}). These expressions can be found in the Appendix.

\subsection{The model of nuclear structure\label{sec:nucl}}

The matrix elements in Eq.~(\ref{gamma}) require the definition of the 
initial and final states of the system. We are dealing with closed--shell
nuclei and then our initial state is built by coupling the $\pi^-$ atomic
wave function with the closed--shell core. Thus the
initial state is given by: 
\begin{equation}\label{eq:pi-state}
|i\rangle \equiv | J_i M_i \rangle \, = \, 
\left[ |\pi^- ; n l  \rangle \otimes |0^+ \rangle \right]^{J_i}_{M_i}
\end{equation}
with $J_i = l$ and  $|0^+\rangle$ the Slater determinant 
corresponding to the closed--shell
nucleus. The Slater determinant is built 
with single--particle wave functions obtained
by diagonalizing a Woods--Saxon potential well 
\begin{equation}\label{eq:pot}
   V_{\rm Woods-Saxon} (r)  
=  V_0 f(r,R_0,a_0) +  V_{LS}
   \frac{{\bf l}\cdot\mbox{\boldmath $\sigma$}}{r}
   \frac{df(r,R_{LS},a_{LS})}{dr}+ 
    V_C(r),
\end{equation}
where
\begin{equation}
f(r,R,a) =  \frac{1}{1+\exp \left(\frac{r-R}{a} \right)}
\end{equation}
and $V_C(r)$ is the Coulomb potential created by an homogeneous
charge distribution of radius $R_C$. The parameters of this potential 
are adjusted to reproduce different nuclear properties such as, e.g., 
the experimental single--particle energies around the Fermi level, 
the r.m.s. radius or
the charge density~\cite{Ama94b}. On the other hand the pion state, 
$|\pi^- ; nlm_l \rangle $,  
is defined as follows:
\begin{equation}\label{eq:pion-field}
\phi_-({\bf r}) \, |\pi^- ; nlm_l 
\rangle \, = \, \frac{1}{\sqrt{2E^\pi_{nl}}} 
R^{\pi}_{nl}(r) \, Y_{lm_l}({\bf \hat{r}}) |0\rangle_\pi
\end{equation}
with $|0\rangle_\pi$ the pion vacuum state,  $n,l$ and $m$ 
the principal, orbital angular momentum and magnetic  quantum numbers of
the pion atomic orbit and $Y_{lm}({\bf \hat r})$  
spherical harmonic functions. 
The radial pion wave function $R^{\pi}_{nl}(r)$
and the pion energy $E^\pi_{nl}$ are obtained by solving
the Klein-Gordon equation with a pion-nucleus optical potential. This
potential has been developed microscopically and it is exposed in detail
in Ref.~\cite{NOG93} for pionic atoms. It contains the coulombian $\pi^-$-
nucleus interaction and the ordinary lowest
order optical potential constructed from the $s$-- and $p$--wave $\pi N$
amplitudes. In addition second order terms in both $s$-- and $p$--waves, 
responsible for pion absorption, are also considered. 
Standard corrections to both the electromagnetic (finite size of  the 
nucleus and vacuum polarization)
and strong  (second-order Pauli re-scattering term, ATT term,
Lorentz--Lorenz effect, long and short range nuclear correlations)
parts of the potential are also taken into account. This theoretical potential
reproduces fairly well the data  of pionic atoms (binding energies and
strong absorption widths)~\cite{NOG93} and  low energy $\pi$-nucleus scattering
~\cite{NOG93bis}.

In Eq.~(\ref{gamma}), the initial energy $E_i$ corresponds 
to the energy of the $(\pi X)_{\rm bound}$ system, 
including the pion binding energy. As we are only interested
in the excitation nuclear energy, we take the energy of the
initial closed--shell nucleus as our zero of energies. Thus the
initial energy $E_i$ can be identified as the pion mass plus its
binding energy, $E^\pi_{nl}$.

The final states are described in terms of a 
neutron particle in the
continuum and a proton hole coupled to the final angular momentum
$J_f$, 
\begin{equation}\label{eq:ph}
| f \rangle \equiv |ph^{-1}; J_f M_f \rangle = 
\left[  a^{\dagger}_p\otimes  b^{\dagger}_h 
\right]^{J_f}_{M_f} \, |0^+\rangle
\end{equation}
where $a^{\dagger}_p$ ($b^{\dagger}_h$)
 is the particle (hole) creation operator. The particle
states are obtained by solving the Schr\"odinger equation for positive 
energies. We have used the same Woods--Saxon potential as for the discrete
single--particle levels in order to ensure the orthogonality of the states
involved in the process (see Ref.~\cite{Ama93} for details). It is
important to mention here that, in view of the definition of the final
states we are using, the sum over final states $| f \rangle$ in 
Eq.~(\ref{gamma}) should be changed to:
\begin{equation}\label{eq:sum}
\sum_{f} \, \longrightarrow \, \sum_{ph} \,\sum_{J_f}.
\end{equation}

Therefore,  to evaluate Eq. (\ref{gamma}) 
we must perform a sum over the quantum numbers of the final particle-hole
states ($h =n_h,l_h,j_h$, $p=\epsilon_p, l_p, j_p$ and the total 
angular momentum $J_f$ of the $ph$ excitation) 
and the photon angular momentum ($J$). The sum over
$\epsilon_p$ is indeed an integration which is performed by means of
the delta of energies which appears in Eq. (\ref{gamma}). Actually, in
our model, the final energy $E_f$ is given by
\begin{equation}
E_f \, = \Delta m + \, \epsilon_p \, - \, \epsilon_h ,
\end{equation} 
with $\epsilon_\alpha$ the energy of the single--level $\alpha$ and
$\Delta m$ the difference between the masses of the particle (neutron)
and hole (proton) nucleons. Thus, for
each value of the photon energy $q$ and each hole $h$, 
$\epsilon_p$ is univocally 
determined by the delta function in Eq.~(\ref{gamma}).

Note also
that the sum over the photon multipoles, $J$, ranges from one to infinity. 
The larger is the photon energy or the nuclear mass, the bigger is
the number of photon multipoles needed to obtain convergence.  In the
RPC process, the maximum photon energy is around the pion mass and we
have found in $^{40}$Ca a good convergence with four or five
multipoles.

One final detail concerns to the hole-energies used in our
calculation. As it was mentioned above, our
mean-field potential has been adjusted to reproduce single-particle
energies around the Fermi level. To improve the energetic balance in 
our model, in the actual calculation the experimental 
proton single-particle energies values, when known,  have been used. 
In Table~\ref{tab:single-energies}, we compare these values to 
the computed mean-field proton hole energies.

\begin{table}
\begin{center}
\begin{tabular}{ccll}\hline
Nucleus  & Shell & $\epsilon_h (th)$ & $\epsilon_h$ (exp) \\ \hline
$^{12}$C & $1p_{3/2}$  & -18.05             &  -15.96  \\\hline
$^{16}$O & $1p_{1/2} $  & -12.74 &   -12.12     \\
         & $1p_{3/2} $  & -16.92 &   -18.45         \\\hline 
$^{40}$Ca& $2s_{1/2}$& -10.30   &  -10.85            \\
         & $1d_{3/2}$& -8.71    &  -8.33              \\
         & $1d_{5/2}$& -16.19   &  -12.27             \\\hline
\end{tabular}
\end{center}
\caption{Proton hole single-particle energies (MeV). We show the
theoretical energies computed with the SM potential described
in sect.~\protect\ref{sec:nucl} (third column) and the experimental ones 
(fourth column). For more details, see Ref.~\protect\cite{Ama94b}. } 
\label{tab:single-energies}
\end{table}

\subsection{Effects of the $2p2h$ correlations.}\label{sec:2p2h}

In sect. \ref{sec:nucl}, we
have outlined our continuum SM approach to the nuclear 
structure. In this framework, the 
 final state interactions (FSI) of the outgoing particle
(neutron) with the residual nucleus are partially taken into
account. However, $2p2h$ correlations are a fundamental ingredient to
achieve a reasonable descriptions of these FSI. Here, we discuss
briefly the phenomenological model  for such effects used in this work.
We follow the approach of Ref.~\cite{Smi88}. There, the propagation of
the mean field $ph$ excitations in the nuclear medium is modified by
including an approximated complex $ph$ self--energy. 
This self--energy is determined by
the effective nucleon-nucleon residual interaction in the medium not
included in the mean field approach.  The inclusion of the $2p2h$ 
correlations improves our treatment of the FSI and it also accounts 
for corrections to the pure one--body RPC due to two nucleons mechanisms. 
This approach has  proved to be
successful in the study of the transverse~\cite{Ama94}  and  
longitudinal~\cite{Co88} 
response functions in the quasielastic region for inclusive 
$(e,e^\prime)$ processes. Also a  recent approach~\cite{Gil97},
 where the propagation of $ph$ excitations in the medium is described 
in terms of particle and hole spectral functions 
gives similar results to those of Refs.~\cite{Ama94} and~\cite{Co88}.

In previous approaches~\cite{Ver75}, FSI were
taken into account by putting the outgoing neutron in a wave of an
neutron-nucleus optical  potential. The imaginary part of the optical
potential reduces the outgoing neutron flux and therefore reduces the 
integrated RPC rate. However, we believe this procedure is 
incomplete: the imaginary part of the optical potential  is mainly due to 
two--body mechanisms which
also contribute to the RPC rate and have to be considered in order to
be consistent (see Ref.~\cite{Chi90} for more details). 
Our approach, where we incorporate a
$ph$ self--energy, automatically takes into account both physical
processes mentioned above, leading to a redistribution of the 
strength of the differential photon rate  but
as we shall see leaving the integrated RPC rate almost unchanged.

Following the steps of Ref.~\cite{Co88} to include the $2p2h$
correlation effects to the RPC differential rate, one should fold the 
pure one body SM differential rate, 
$\left ( d\Gamma^{(\gamma)}_{nl} (q)/dq \right )^0 $, given in
Eq. (\ref{gamma}), with the electromagnetic current and
final states defined in Eqs.(\ref{current}) and (\ref{eq:ph}), with
the imaginary part of the $ph$ propagator:
\begin{eqnarray}
\frac{d\Gamma^{(\gamma)}_{nl} (q)}{dq} & = & \int^{m_\pi}_0 dq\,' 
\left ( \frac{d\Gamma^{(\gamma)}_{nl} (q\,')}{dq\,'} \right )^0 \left
[\rho(m_\pi-q\,',m_\pi-q) + 
\rho(m_\pi-q\,',q-m_\pi) \right ] 
\end{eqnarray}
with
\begin{eqnarray}
\rho(E,\omega) & = & \frac{1}{2\pi}\frac{\Gamma^{\downarrow}(\omega)}
{(E-\omega-\Delta(\omega))^2+(\Gamma^{\downarrow}(\omega)/2)^2}.
\end{eqnarray}
The functions $\Delta(\omega)$  and  $\Gamma^{\downarrow}(\omega)$ define 
the complex self--energy,
\begin{equation}
\Sigma (\omega) =   \Delta(\omega) - \frac{i}{2}\Gamma^{\downarrow}(\omega)
\end{equation}
of a $ph$ with excitation energy $\omega$. The self--energy $\Sigma
(\omega)$  has been taken from Ref.~\cite{Co88}.

\section{Results} 

In this section we apply our model to calculate the RPC differential rates
for three closed--shell medium nuclei, $^{12}$C, $^{16}$O and $^{40}$Ca, and
compare our predictions with the available experimental data.

Experimentally, it is rather difficult to distinguish between RPC 
processes from different pionic atom orbits. Indeed, only the
weighted ratio 
\begin{eqnarray}\label{eq:def-R}
\frac{dR^{(\gamma)}}{dq} = \sum_{nl}
\frac{\omega_{nl}}{\Gamma^{abs}_{nl}}
\frac{d\Gamma^{(\gamma)}_{nl}}{dq}
\end{eqnarray}
can be measured.  Here,  $\omega_{nl}$ gives the absorption
probability  from each $nl$
pionic level, taken into account  the electromagnetic transitions and the
strong absorption, and are normalized to the
unity. $\Gamma^{abs}_{nl}$ is the total pion absorption width from the orbit
$nl$ and $\Gamma^{(\gamma)}_{nl}$ is the width 
due to the radiative capture of the pion from the orbit $nl$. As
disccused in~\protect\cite{Bae77}, we make the
approximation of putting all the weight of radiative capture in the
two pionic levels observed with the X-ray technique. The values used
in this work for  these parameters are shown in Table
\ref{tab:wnl-gabs}. In what follows we will
refer to $dR^{(\gamma)}/dq$  instead of the single differential decay rate
of Eq.~(\ref{gamma}).

\begin{table}
\begin{center}
\begin{tabular}{cccl}\hline
Nucleus  & $nl$ & $\omega_{nl}$ & $\Gamma^{abs}_{nl}$ (KeV) \\ \hline
$^{12}$C &  1$s$    &     0.1          &   3.14               \\
         &  2$p$    &     0.9          &   0.00136               \\ \hline
$^{16}$O &  1$s$    &     0.1	       &   7.92		   \\
         &  2$p$    &     0.9          &   0.00676         \\ \hline
$^{40}$Ca&  2$p$    &	  0.7	       &   1.59		\\
         &  3$d$    &     0.3          &   0.0007        \\ \hline
\end{tabular}
\end{center}
\caption{Values of $\omega_{nl}$ \protect\cite{Bae77} and 
$\Gamma^{abs}_{nl}$ \protect\cite{Bac70}---\protect\cite{Oli85} used in this
work.} 
\label{tab:wnl-gabs}
\end{table}

\begin{figure}

\vspace{-5cm}

\begin{center}                                                                
\leavevmode
\epsfysize = 750pt
\makebox[0cm]{\epsfbox{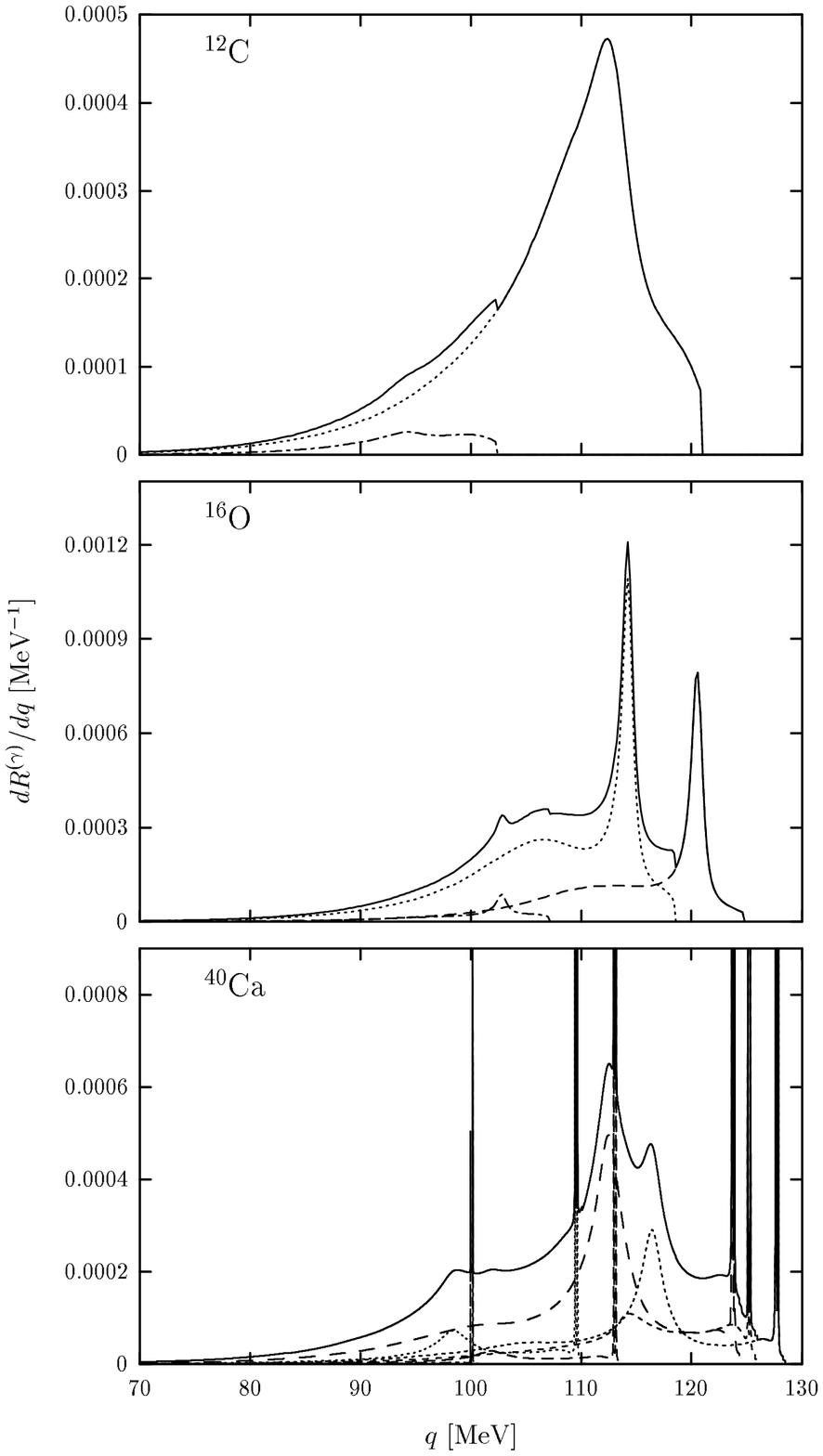}}
\end{center}
\vspace{-5.5cm}
\caption[]{Contribution to $dR^{(\gamma)}/dq$ from the
different holes configurations in $^{12}$C, $^{16}$O and
$^{40}$Ca. In all figures, the solid lines stand for the total sum
of the contribution of the hole configurations. 
Each hole can contribute up to a maximum 
photon energy, $q^{(h)}_{max} = E_i-\Delta
m + \epsilon_h$, which is different for each shell. This fact can be used to
identify the contribution of each hole in the figures: the deeper is
the shell, the smaller is $q^{(h)}_{max}$. From the deepest to the outest
shell, the hole configurations involved are: $1s_{1/2}$, $1p_{3/2}$, 
$1p_{1/2}$, $1d_{5/2}$, $2s_{1/2}$ and  $1d_{3/2}$.} 
\label{fig:fig1}
\end{figure}

We start now discussing our results. Firstly, we focus on the purely
one-body process and later we will introduce the effects of the 
$2p2h$ correlations. In Fig.~\ref{fig:fig1}, 
we show the contribution to
$dR^{(\gamma)}/dq$ from the different holes configurations 
in  $^{12}$C, $^{16}$O and $^{40}$Ca. The most important contributions 
are due to the $1p$ shell in  $^{12}$C  and  $^{16}$O  and to the 
$1d$ and $2s$ shells in  $^{40}$Ca. In all cases they correspond to the
outest hole configurations, which maximize the overlap with the
pion wave function. The end of the contribution of a particular shell
produces a protuberance in the total sum. On the other hand, in the
plots also appear resonance structures, specially in 
$^{40}$Ca where turn out to be extremely narrow with typical 
widths of the order of 10 KeV. This was already pointed out
in~\cite{Co87}. However, the resonance structure for low nuclear
excitation energies  depends strongly on the details of the potential. 
The use of a different nuclear
structure model will produce drastic changes of both their 
widths and positions. 

Note also that in the data,  
these resonances will be convoluted with the
experimental photon energy resolution, which is much greater than
their  widths and therefore the sharp resonances shapes will smear out
and a smoother spectrum will be observed. Here, we are mainly 
interested in the global shape of the photon distribution and the
study of these fine details is out of the scope of this work, because
it will require both a more refined nuclear  model  
and more precise experimental data.

\begin{figure}

\vspace{-5cm}

\begin{center}                                                                
\leavevmode
\epsfysize = 750pt
\makebox[0cm]{\epsfbox{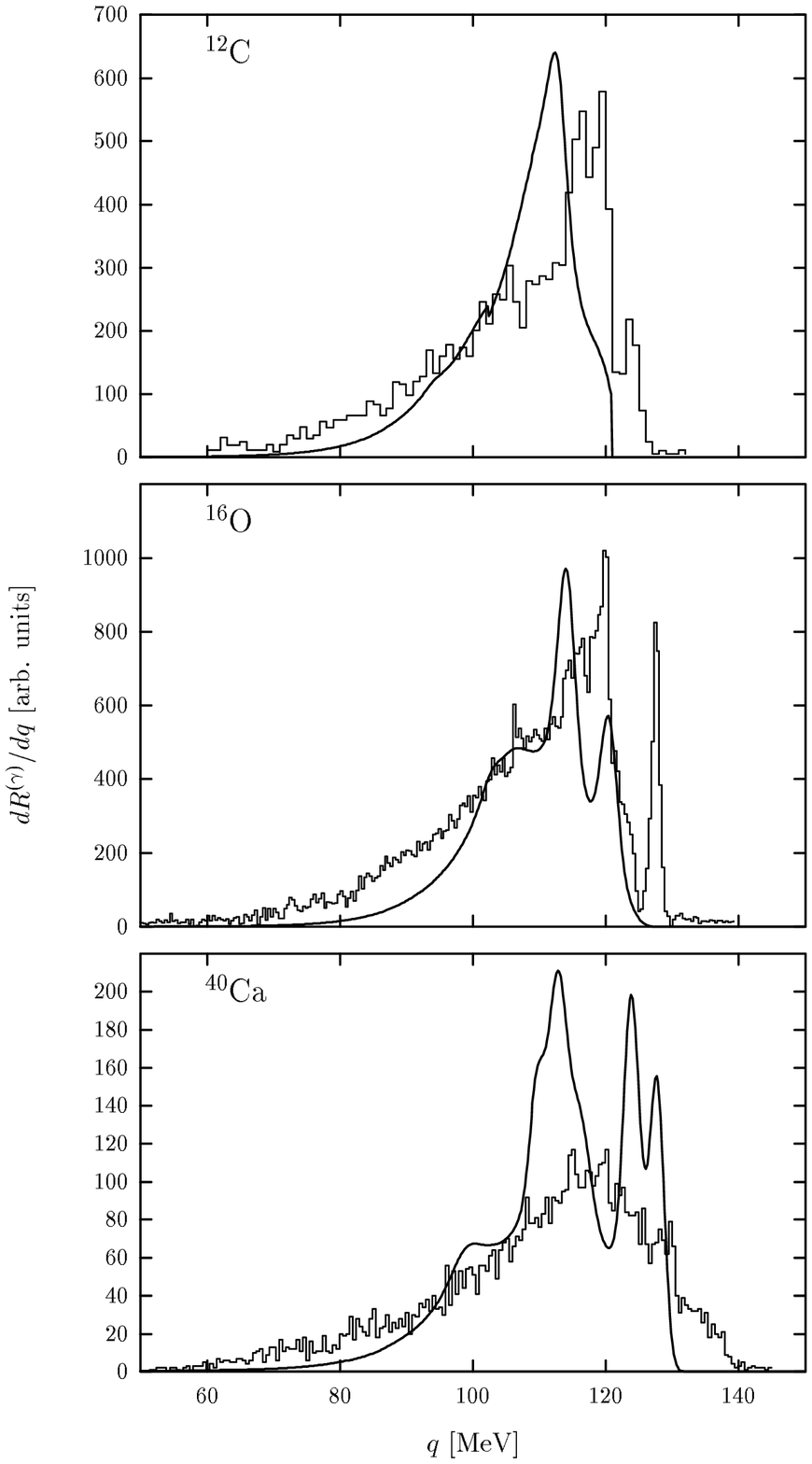}}
\end{center}
\vspace{-5.5cm}

\caption[]{Results for $dR^{(\gamma)}/dq$ in 
$^{12}$C, $^{16}$O and $^{40}$Ca. Only one-body contributions have
been included. Experimental data, taken from~\protect\cite{Bae77} and
\protect\cite{Bis72}, are given in arbitrary
units. Our results have been normalized to the data at $q=105$ MeV.
In oxigen and calcium we have convoluted our differential decay rate
with a Gaussian weight function of width $1.5$ MeV. } 
\label{fig:fig2}
\end{figure}

In Fig.~\ref{fig:fig2} we compare the total sum
of all hole contributions shown in  Fig.~\ref{fig:fig1} with the 
experimental data.  In calcium and oxigen, following the discussion
above, we have eliminated the sharp
resonance structures in our differential rates by convoluting the
responses with a Gaussian weight function $f(\omega) \propto
exp(-\omega^2/\Gamma^2)$ of width $\Gamma=1.5$MeV. This convolution
procedure keeps unchanged the integrated RPC rate~\cite{Ama96}. The 
experimental data are given in arbitrary units, 
thus to compare with the experiment
we have normalized our results to the data at $q=105$ MeV. The reason
to normalize our results in this region of intermediate energies is 
twofold. Firstly because is far from the low nuclear excitation energy region
(high photon energies) where, as we discuss above, the theoretical
results depend strongly on the nuclear model. Secondly, because the
low photon energy region is sensitive to the high momentum components
of the hole wave function, and one expects theoretical 
uncertainties in the nuclear model. Furthermore, as we will show below, 
the effect of $2p2h$ correlations should be also included in the low energy 
photon region. 

As it can be seen in Fig.~\ref{fig:fig2} our
description gives the gross features of the experimental
data. However, at the lower end of the photon
spectrum the tail of the theoretical distribution is 
significantly narrower than the experimental one. Furthermore, at high
photon energies, major discrepancies also appear because the contributions
there  correspond to processes where the outgoing neutron is either 
bound in the residual nucleus or it is in the continuum but carrying
low kinetic energies. The first contributions have not been 
considered in our approach yet while the latter ones depend strongly
on the details of the nuclear structure model.

\begin{figure}

\vspace{-3cm}

\begin{center}                                                                
\leavevmode
\epsfysize = 850pt
\makebox[0cm]{\epsfbox{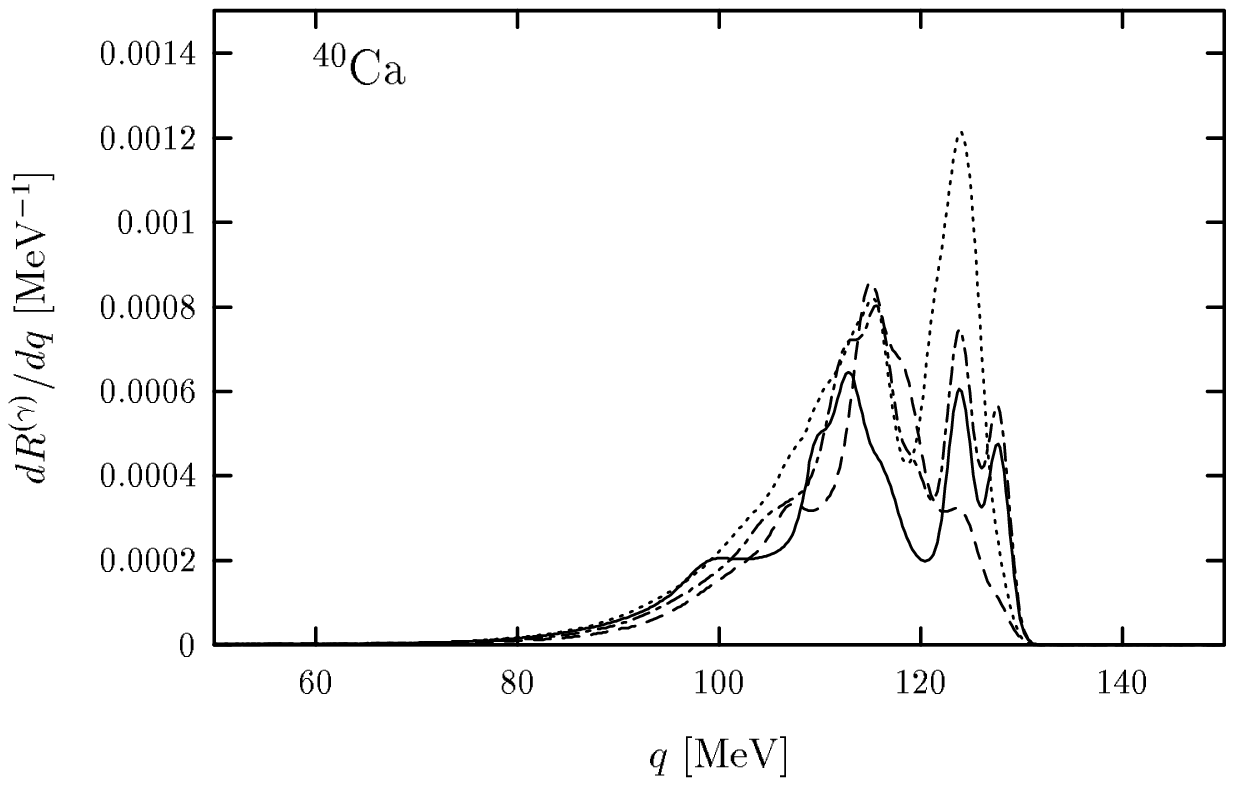}}
\end{center}
\vspace{-19.5cm}

\caption[]{Results for $dR^{(\gamma)}/dq$ in 
$^{40}$Ca with the different SM potentials defined
in Table~\protect\ref{tab:pot}. The solid line
corresponds to the potential A outlined in
sect.~\protect\ref{sec:nucl} and used in the rest of this paper. 
Dash-dotted, dotted and dashed curves correspond to the potentials 
B,C and D respectively.} 
\label{fig:fig3}
\end{figure}

One may think that the situation could be improved by allowing for 
reasonable modifications of the mean field potential. 
In Fig.\ref{fig:fig3} we show  the results for  $dR^{(\gamma)}/dq$
in $^{40}$Ca obtained with the four different potentials defined in 
Table~\ref{tab:pot}. They differ each other only in the proton or 
neutron depth of the Wood-Saxon well (parameter $V_0$ 
in Eq. (\ref{eq:pot})). We see in the figure that none of the new
potentials (B,C,D) give a photon distribution significantly wider than
the A potential, used throughout this work. However, 
the behavior above  $110$ MeV (low nuclear
excitation energies) turns out to be very sensitive to the potential.
The area below the curves depends also appreciably on the model and 
it increases when the neutron Wood--Saxon depth decreases.

\begin{table}
\begin{center}
\begin{tabular}{cccc}\hline
Potential  & $V_0$ (protons) & $V_0$ (neutrons) & $R^{(\gamma)} [\%]$ \\
\hline
  A        & $-57.5$       & $-55.0$   &  $1.25$              \\
  B        & $-50.0$       & $-55.0$   &  $1.54$             \\
  C        & $-50.0$       & $-50.0$   &  $1.87$      \\
  D        & $-50.0$       & $-60.0$   &  $1.21$
\\\hline 
\end{tabular}
\end{center}
\caption{ Different Woods--Saxon potentials considered in this work. 
 The second and third columns give proton and neutron 
 Wood--Saxon well depth 
parameters (in MeV) for $^{40}$Ca 
used to compute the different curves appearing 
in Fig.~\protect\ref{fig:fig3}. The rest of the parameters, not shown
here, are common for all potentials  and can be found 
in Ref.~\protect\cite{Ama94b}. The potential B has been adjusted to reproduce
the experimental proton density, whereas C(D) corresponds to a 10\%
decrease (increase) of the depth for the neutron Wood-Saxon
well. The fourth column gives the integrated ratio $R^{(\gamma)}$
(in percent) for each potential. } 
\label{tab:pot}
\end{table}

\begin{figure}

\vspace{-5cm}

\begin{center}                                                                
\leavevmode
\epsfysize = 750pt
\makebox[0cm]{\epsfbox{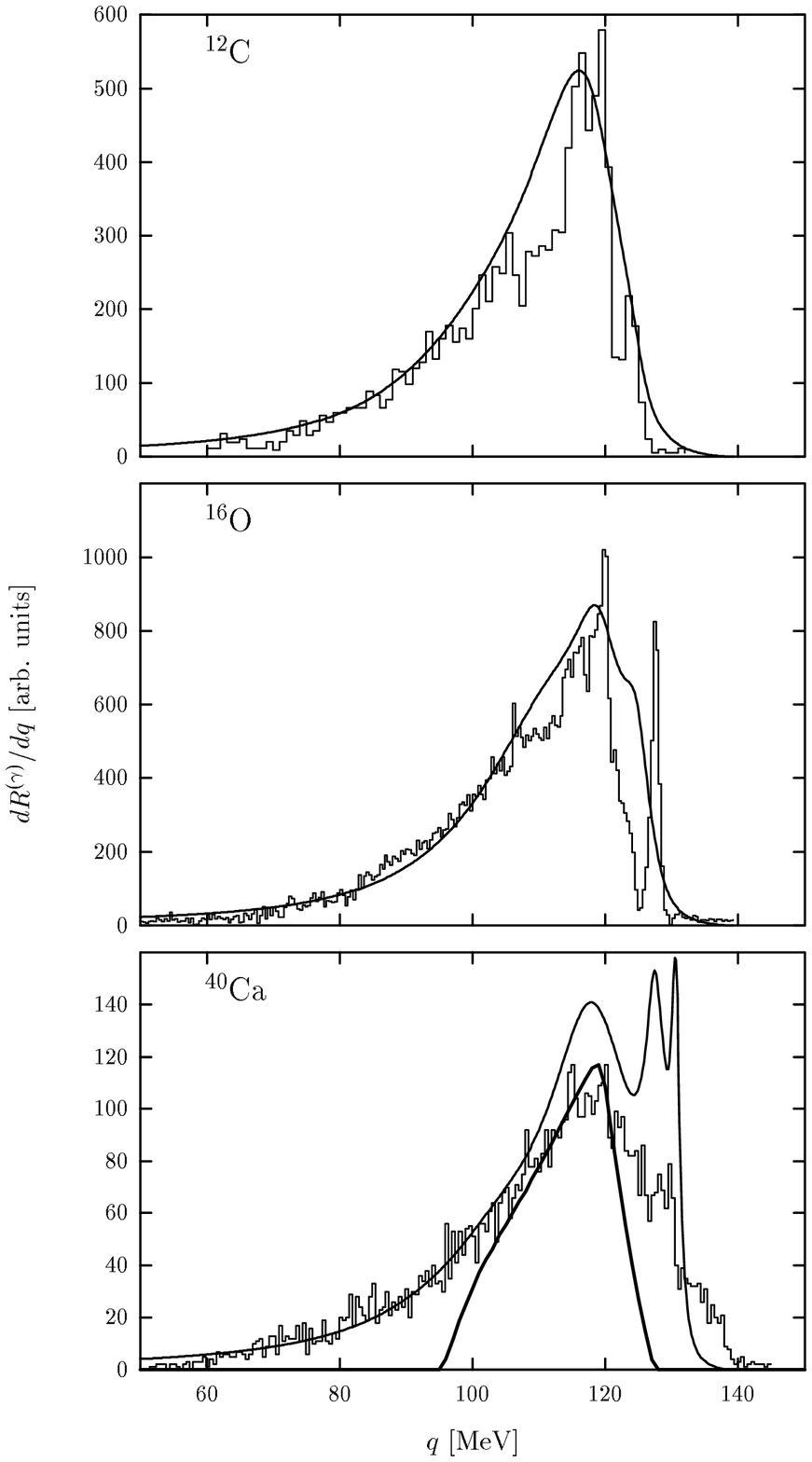}}
\end{center}
\vspace{-5.5cm}

\caption[]{Results for $dR^{(\gamma)}/dq$ in 
$^{12}$C, $^{16}$O and $^{40}$Ca. Two-body contributions have
been now included. Experimental data, taken from~\protect\cite{Bae77} 
and \protect\cite{Bis72}, are given in arbitrary
units. Our results have been normalized to the data at $q=105$
MeV. For comparison in $^{40}$Ca we have also shown the FG 
results presented in Fig. (3) of Ref.~\protect\cite{Chi90} (thick
line). The absolute value of the FG distribution around the
peak is about a factor two or three greater than the SM one 
(thin line) and it has been normalized to the peak of the
data. Additionally, the FG curve has been shifted 10 MeV to the
left.}
\label{fig:fig4}
\end{figure}

Now we focus our attention in the effects of the $2p2h$
correlations. Following the method described in 
sect.~\ref{sec:2p2h}, we include the effects of the 
$ph$ self--energy in the medium,  into the pure one--body SM
differential rates of Fig.~\ref{fig:fig1}. 
The results are shown in Fig.~\ref{fig:fig4}. We find an excellent
agreement with the experimental distribution, specially in the low
energy tail. Clearly, the inclusion of 
two body mechanisms has made the distribution wider and turns out to
be essential to explain the tail of the experimental data. 
For comparison  we also show the results for  $^{40}$Ca with the model 
of Ref.~\cite{Chi90}. In this reference, the calculations were 
performed  in nuclear matter and the LDA 
was used to obtain results in finite nuclei. There, though it was found a fair
agreement for the integrated ratio $R^{(\gamma)}$ throughout the
periodic table, it was pointed out the inability of the model to
describe the tail of the photon distribution. In Ref.~\cite{Chi90}
 this problem was
associated to the lack of high-momentum components in the
FG nucleon wave functions. Here, we show that, 
in addition to the use of proper nucleon wave functions, one needs to
incorporate $2p2h$ correlations  to understand the experimental distribution.

\begin{figure}

\vspace{-3cm}

\begin{center}                                                                
\leavevmode
\epsfysize = 850pt
\makebox[0cm]{\epsfbox{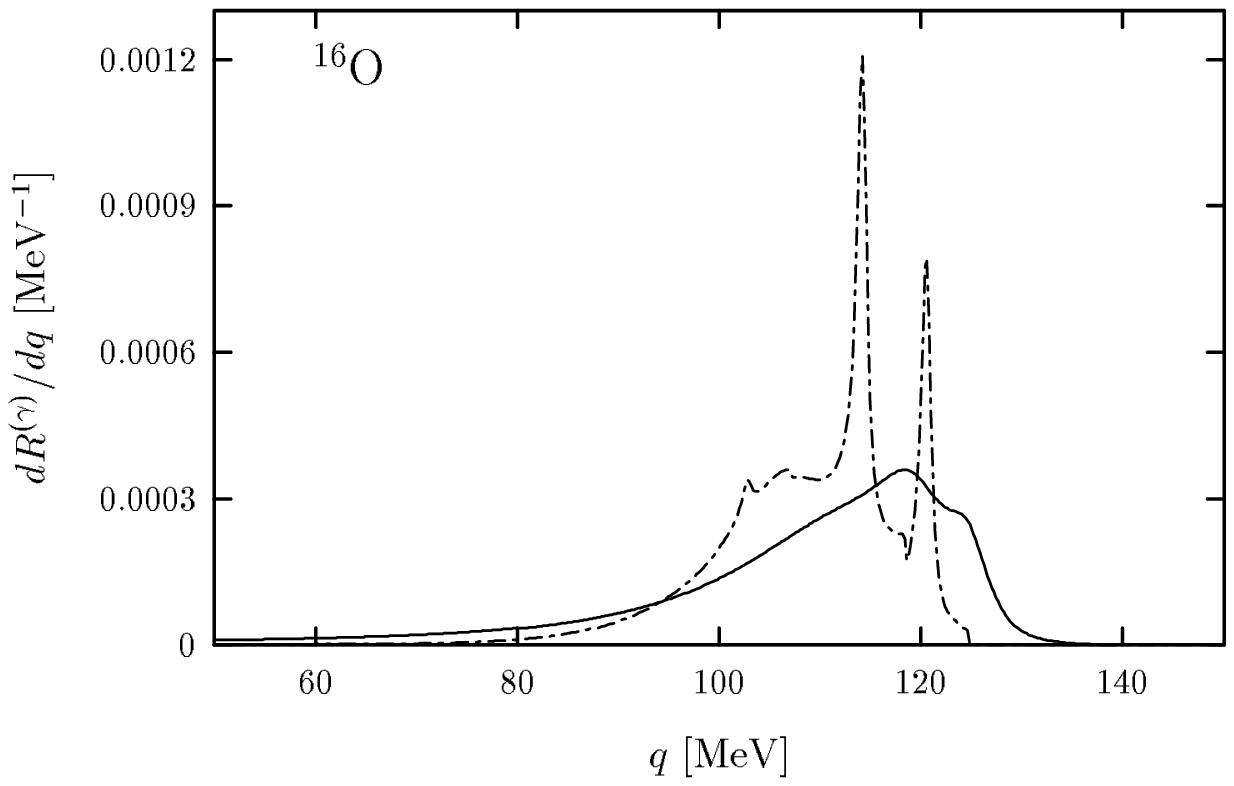}}
\end{center}
\vspace{-19.5cm}

\caption[]{Results for $dR^{(\gamma)}/dq$ in 
$^{16}$O with (solid line) and without (dashed line) the inclusion of
$2p2h$ correlation effects. The area below the curves are $9.2
\times 10^{-3}$ and $9.8 \times 10^{-3}$ respectively.} 
\label{fig:fig5}
\end{figure}

To discuss in more detail the effect of the inclusion of
$ph$ self--energy in the calculation of the differential decay rate,
we show in Fig.~\ref{fig:fig5} the results for $dR^{(\gamma)}/dq$ in  
$^{16}$O with and without the inclusion of $2p2h$ correlations. 
As it can be seen, these effects reduce the
strength of the differential rate in the region of energies around the peak 
due to the pure one--body absorption, 
as it would do a FSI treatment with an optical
potential. But, in addition these $2p2h$ correlations incorporate 
the contribution to the RPC process due to two--nucleons
mechanisms in the low energy part of the photon spectrum. The net
effect is a redistribution of the strength in the spectrum, 
with a variation of less than 10\% in the integrated rate 
$R^{(\gamma)}$. Similar effects have
been also found in inclusive $(e,e')$ processes  at the quasielastic
~\cite{Ama93},~\cite{Co88}, \cite{Gil97} and
$\Delta$ peaks~\cite{Gil97} and also in inclusive $(p,p')$ and $(n,p)$ 
processes~\cite{Smi88}.

Besides, we would like to stress that the use of the same mean
field potential for particle and hole states in our calculations,
guarantees the orthogonality of the $ph$ states. 
It is common in the literature 
(\cite{Ver75}, \cite{Wun80}) to use different
potentials for particles and holes (optical models for the particles
and mean field potentials for the holes, etc...) and therefore the
orthogonality between particle and holes states is lost. In these
circumstances, there are new contributions to the transition matrix
elements coming from the overlap between non-orthogonal
wave functions. These new contributions, though not significant for high
momentum transfer processes~\cite{Bof82}, 
can be important for processes like this
where both energy and momentum transfered to the nucleus are small.

The last aspect we want to discuss concerns the integrated 
ratio $R^{(\gamma)}$. In Table~\ref{tab:resultados-R} 
we present our results for this ratio  and compare them with experiment
and with those of the FG model of Ref.~\cite{Chi90}.

\begin{table}
\begin{center}
\begin{tabular}{cccccc}\hline
& \multicolumn{5}{c}{$R^{(\gamma)}$ [\%]} \\\hline
  &
Ref.~\protect\cite{Chi90}& Continuum & Discrete & Total & Experiment \\ \hline
$^{12}$C & 
$1.19 \pm 0.16$ &0.62 & 1.17& 1.79 & $1.84 \pm 0.08$, $1.92 \pm 0.19$,
$1.6 \pm 0.1$      \\\hline
$^{16}$O & $1.53 \pm 0.08$ & 0.92 & 0.80& 1.72 &  $2.27 
\pm 0.24$, $2.24 \pm 0.48$\\\hline 
$^{40}$Ca&  $1.91 \pm 0.29$ & 1.17 & 0.87 & 2.04 & $1.82 \pm 0.05$ \\\hline
\end{tabular}
\end{center}
\caption{Results (in percent) for $R^{(\gamma)}(q)$ in 
$^{12}$C, $^{16}$O and $^{40}$Ca. The second column contains 
the results obtained with the FG model of
Ref.~\protect\cite{Chi90}. Our results are given in the next three
columns: third column corresponds to the area below the curves in 
Fig.~\protect\ref{fig:fig4}, fourth column gives the
contribution of the RPC processes 
in which the final neutron is bound in
the residual nucleus and finally the fifth column is the total rate,
sum of the two previous values. Experimental values are taken from references:
\protect\cite{Bis72}, \protect\cite{Dav66} and \protect\cite{Pet65}  
 for $^{12}$C, \protect\cite{Bis72} and \protect\cite{Str79}  
 for $^{16}$O, and \protect\cite{Bis72} for $^{40}$Ca.} 
\label{tab:resultados-R}
\end{table}

The results obtained within the framework described up to here (column
labeled ``Continuum'' in the table)  turn out to be around a factor
of two smaller than the experimental ones and those of 
Ref.~\cite{Chi90}. However, it is worth to
point out here that, as shown in Table~\ref{tab:pot}, the integrated rate  
is rather sensitive to the details of the SM nuclear
potential used, and that a small variation of the model 
could account for such discrepancies. Apart from this point, 
other possible reasons for the low rates
found  are the following:

\begin{enumerate}

\item  We have only considered the 
Kroll--Ruderman term for the elementary process $\pi^- p \to \gamma
n$. The momentum of the pion on its bound orbit though small is not 
strictly zero, and one should add to the dominant Kroll--Ruderman term
new amplitudes proportional to the pion momentum which could increase
the integrate RPC rate. In any case these new contributions are not
expected to be important enough to explain the disagreement.

\item The determination of the weights $\omega_{nl}$  
(see Table~\ref{tab:wnl-gabs}) may be subject to
theoretical uncertainties. It would be interesting
to disentangle experimentally the capture from different pionic 
atomic states to allow a direct comparison with the theory, free of
the assumptions made on the values of $\omega_{nl}$ (for more details
see Ref.~\cite{Bae77}).

\item Pion capture processes where the outgoing neutron is
bound in the residual nucleus instead of going to the continuum lead
to Dirac's delta peaks in the photon spectrum which also
contribute to the integrated ratio $R^{(\gamma)}$ and  have
not been considered yet. Indeed, one can easily evaluate 
these contributions by taking into account that  
the integral over $\epsilon_p$ in Eq.~(\ref{eq:sum})  is now a
discrete sum over the unoccupied single neutron states (see
Ref.~\cite{Ama94b})  and  the radial wave function $R_p$ in 
Eq.~(\ref{eq:rp-rh}) corresponds now to the solution of the radial 
Schr\"odinger equation for a negative energy $\epsilon_p$.

In the fourth column of Table~\ref{tab:resultados-R} we give the
contributions from these discrete states to the integrated rate. We see
that they are of the same order of magnitude than the
continuum ones and lead to a reasonable agreement with the
experimental data. To describe properly the magnitude and the position
of each individual discrete transition, one would need a more sophisticated
model for the residual $^{12}$B, $^{16}$N and $^{40}$K nuclei than the
one used here, as it was the case when we discussed the continuum
integrated rates.  However,  the sum of all the discrete and continuum
contributions is rather model independent as it can be seen in 
Table~\ref{tab:pot-2}. Results with potentials B, C and D 
agree within around $4 \%$, while the value obtained with potential
A is well below the previous ones. This is because the proton well 
for potential A is more attractive than for the rest of potentials, 
and therefore the overlap between the holes and pion wave functions 
is smaller. Changes in the  proton potential modifies both the
continuum and discrete contributions, and then the total
rate. However, variations of the neutron potential depth, leaving fixed the
proton one, make some single--particle states to be
bound or not. In this last case, low energy resonances appear in the
spectrum. Thus, the net effect is an exchange of strength 
between the discrete and continuum contributions to the RPC rate,
leaving almost unchanged the total value.

From the above discussion we see that reasonable changes in the proton 
potential lead to about 15-20\% uncertainty in the SM
results. Moreover, one gets an additional 4\% uncertainty associated
to the neutron potential, thus we 
conclude that the SM gives integrated RPC rates with an
uncertainty of about 20-25\%. Similar results are also found for 
carbon and oxigen. This uncertainty is basically due to two facts:
{\it i)} the RPC process takes place at the nuclear surface  and {\it
ii)} the main contribution to the integrated RPC rate comes from nuclear
excitation energies rather low ($\approx 5-10$ MeV). For processes
sensitive to the whole nuclear volume and involving 
larger excitation energies (see for instance calculations at the 
quasielastic peak for inclusive  $(e,e')$ scattering ~\cite{Ama93})
the SM results are much more independent of the details of 
the mean field.

\begin{table}
\begin{center}
\begin{tabular}{ccccc}\hline
& \multicolumn{4}{c}{$R^{(\gamma)}$ [\%]}\\\cline{2-5}
& \multicolumn{3}{c}{Shell--model} & \\\cline{2-4}
Potential/Density  & Continuum & Discrete & Total & LDA Fermi gas \\
\hline
    A        & 1.17 (1.25)       & 0.87 (0.92)   &  2.04 (2.17)  & 2.20 \\
    B        & 1.44 (1.54)      &  0.97 (1.03)   &  2.41 (2.57)  & 2.62  \\
    C        & 1.76 (1.87)      & 0.57 (0.61)   &  2.33  (2.48)  & 2.59  \\
    D        & 1.14 (1.21)      & 1.28 (1.36)   &  2.42  (2.57)  & 2.66  \\
   2pF       &  --              & --            &  --            &2.74
\\\hline 
\end{tabular}
\end{center}
\caption{Comparison of the results (in percent) for $R^{(\gamma)}(q)$ 
in $^{40}$Ca for the same potentials of the
Table~\protect\ref{tab:pot}  and the FG model of
Ref.~\protect\cite{Chi90}. 
In the first column we give
either the Wood--Saxon potential used to compute the SM
results or the proton and neutron center densities needed as input in 
the FG model. Densities A--D refer to those obtained from the
corresponding Wood--Saxon potentials
A--D and  2pF to the two parameter  Fermi density used in
the Ref.~\protect\cite{Chi90} and taken from
Ref.~\protect\cite{Jag74}. The 2pF value 
does not match the one shown in 
Table~\protect\ref{tab:resultados-R} because here RPA correlations
have not been included and the error bars due to uncertainties in 
the strong absorption widths, $\Gamma^{abs}_{nl}$, have been dropped
out.  To compare with the
FG model we also show in brackets the results of the
SM calculation without taking into account $2p2h$ correlations.} 
\label{tab:pot-2}
\end{table}

\item We have not included  RPA
correlations in our model. In Ref.~\cite{Chi90} RPA lead
to a reduction of about 40\% in the integrated ratios without changing
the shape of the photon distribution.   In
Table~\ref{tab:pot-2} we compare the SM results,
without including $2p2h$ correlations (in brackets), to those obtained
with the LDA FG model of Ref.~\cite{Chi90} without including 
RPA correlations (last row). We see that the FG
model value (2.74) is significantly higher than the SM
ones, part of this discrepancy being due to the use of different proton
and neutron center distributions in both types of calculations. 
To make more reliable the comparison, we also show the LDA FG 
results obtained using the proton and neutron densities derived from
the different Wood-Saxon potentials used in the SM
calculations. We can see now that the FG model gives similar
results to those of the SM and it is also subject to the
same type of uncertainties (20-25\%) associated to the precise 
details of the nuclear dynamics. Similar results are obtained in 
carbon and oxygen. 

However, we would like to point out that the LDA FG results are
in most cases above the SM ones. This can be due to the
undoubted deficiencies of the FG picture of the nucleus to describe 
processes where low nuclear excitation energies are involved. But one 
could also try to explain this disagreement (or part of it) by 
questioning the validity of the LDA  to describe  a surface
process like the one studied here. In any case, because of the zero range
character of  the Kroll--Ruderman interaction considered here, 
LDA should work much better than what one could expect.

In any case, and in view of the results in Table~\ref{tab:pot-2}, 
one might think that the effect of the inclusion of the RPA
correlations could be similar in both models. The fact that the 
SM result shows an uncertainty comparable to the expected 
RPA effects prevents us to  try to include such correlations before 
having a more reliable nuclear model.

\end{enumerate}

\section{Summary and conclusions}
 
In this paper we have reanalyzed the RPC process in light and medium
nuclei.  We have presented a simple continuum SM which
describes successfully the shape of the photon distribution. We have
paid a special attention to the study of the low energy part of the
photon spectrum which could not be fully understood within the different
models found in the literature and in special in the 
model of Ref.~\cite{Chi90}, despite of being the only one able 
to describe the experimental integrated ratios $R^{(\gamma)}$ thorough the
periodic table. We have
found that the effect of $2p2h$ correlations play an essential 
role to describe the tail at low energies of the photon distribution 
and that these effects together with the use of nucleon 
wave functions with non-vanishing high momentum components  
allow us for a correct description of this part of the spectrum. 

We have also
found that the region of high photon energies depends strongly on the 
details of the nuclear model used. In that region when the outgoing
neutron is bound, the photon spectrum exhibits peaks 
characteristic of the energy spectrum of the final nucleus. Also in
this region, when the emitted neutron goes to the continuum, 
there are  peaks corresponding to resonances of the final residual
nucleus. Much effort have been invested in the past to extract
information about these nuclear bound and resonances
states~\cite{Ver75}, ~\cite{Bae77} and ~\cite{Wun80}. We think that a
first necessary step to attack this interesting problem is a correct
understanding of the quasifree background, studied here, which
requires the inclusion of two body mechanisms and to deal with
orthogonal particle-hole states. Thus, results presented
in this paper could be used in future to isolated the contribution of
such nuclear states from the experimental data. This would clearly
represent an improvement with respect previous works 
where the subtraction of the quasifree background was performed by 
means of optical model calculations or the non-theoretical founded pole
model (see Ref.~\cite{Bae77}).

Finally, we have obtained values for the integrated ratio
$R^{(\gamma)}$ in good agreement with those found with the FG
model of Ref.\cite{Chi90}. We have also discussed different mechanisms
which could account for the remaining discrepancies with the
experimental data. 
None of them are expected to make narrower the photon
distribution because, as we have mentioned above, the tail of the
distribution is mainly due to $2p2h$ effects and high momentum
components in the nucleon wave functions and these physical effects
will be present in all new contributions to the RPC rate.

\section*{Acknowledgments}
We are indebted to P. Tr\"uol for providing us with the files containing 
part of the experimental data presented in this work. This research 
was supported by DGES under contract PB95-1204 and by the Junta de
Andaluc\'\i a. 

\appendix
\section*{Appendix \\ Matrix elements of the electromagnetic operators}

In this appendix we give the matrix elements needed to calculate the
width of the RPC process as given by Eq.~(\ref{gamma}). As it can be 
seen in Eq.~(\ref{current}), ${\bf J}({\bf r})$ is in our case a 
one--body operator in the nuclear space. 
Thus both $T^{\rm el}_{JM}$ and $T^{\rm mag}_{JM}$ 
operators  are also. We define the auxiliary operators ${\cal U}^{\rm el,mag}$
by integrating the operators $T^{\rm el,mag}$ over the pionic degrees
of freedom and projecting the result onto a well defined angular
momentum basis.
\begin{eqnarray}
{\cal U}^{\rm el,mag}_{JJ_iJ'M'}(q) & \equiv &  
\left [_\pi\langle 0 | T^{\rm el,mag}_J (q)  | \pi^-; J_i \rangle
\right ]_{J'M'}\\
&&\nonumber\\
& = & \sum_{MM_i} \langle JMJ_iM_i | J'M' \rangle
 _\pi\langle 0 | T^{\rm el,mag}_{JM} (q)  | \pi^-; J_i
M_i \rangle 
\end{eqnarray}
\noindent where the pion states $| \pi; J_i \rangle$ and $|0\rangle_\pi$ where
defined in section~\ref{sec:nucl}. The ${\cal U}^{\rm el,mag}_{JJ_iJ'M'}(q)$
operators act now only on nuclear states and have rank $J'M'$. The
reduced matrix elements of the $T^{\rm el,mag}$ operators can 
be related to those of the ${\cal U}^{\rm el,mag}$ operators, 
\begin{eqnarray}
\langle ph^{-1};J_f \| T^{\rm el,mag}_J (q)  \| J_i \rangle
&=& (-1)^{J_i-J-J_f} \langle p \|\, {\cal U}^{\rm el,mag}_{JJ_iJ_f}(q) 
\,\| h \rangle,
\end{eqnarray}
where the particle and hole states with third isospin component  
$t_p$ and $t_h$ respectively are given by:
\begin{eqnarray}\label{eq:rp-rh}
| p \rangle &=& R_p (r,\epsilon_p)  |\frac12 \, l_p ;  j_p m_p \rangle
 |\frac12 , t_p \rangle \\[5mm]
| h \rangle &=& R_h (r, \epsilon_h)  |\frac12 \, l_h ;  j_h m_h \rangle
 |\frac12 , t_h \rangle .
\end{eqnarray}
As we discussed in sect.~\ref{sec:nucl}, the radial
wave functions $R_p$ and $R_h$ are obtained by solving the
Schr\"odinger equation with the potential described in that section, 
for positive ($\epsilon_p$) and negative  ($\epsilon_h$) energies,
respectively. The values of $\epsilon_h$ are determined by the quantum
numbers $n_h$, $l_h$ and $j_h$ of the occupied shells in the parent
nucleus, whereas the positive energies $\epsilon_p$ are determined by
 energy conservation. The particle radial wave functions $R_p$ are
normalized to the Dirac's delta of energies.

By using Eqs.~(\ref{electric})-(\ref{current}) and 
Eq.~(\ref{eq:pion-field})  one can find the following
expressions for the single-particle 
${\cal U}^{\rm el,mag}_{JJ_iJ'M'}(q)$ operators:
\begin{eqnarray}
{\cal U}^{\rm mag}_{JJ_iJ_fM_f}(q) &= &  i \sqrt{2} \, \frac{f}{m_\pi} \, 
\left( 1 + \frac{m_\pi}{2M} \right) 
\frac{1}{\sqrt{2E^\pi_{n J_i}}} t_- \label{eq:red-u-mag}\\
& & \nonumber \\
& & 
\times j_J(qr)R^\pi_{n J_i}(r) \left [ \,\left [ Y_J (\hat {\bf r}) \otimes 
\sigma \right ]_J \otimes Y_{J_i}(\hat {\bf r})\right ]_{J_fM_f}\nonumber\\
& & \nonumber \\
{\cal U}^{\rm el}_{JJ_iJ_fM_f}(q) &= &   \sqrt{2} \, \frac{f}{m_\pi} \, 
\left( 1 + \frac{m_\pi}{2M} \right) \frac{1}{\sqrt{2E^\pi_{n J_i}}}t_-
\label{eq:red-u-el}\\
& & \nonumber \\
& &  
 \times \frac{1}{\hat J}\sum_{s=\pm 1} s\sqrt{J+\delta_{s,-1}}\,
j_{J+s}(qr)R^\pi_{n J_i}(r) \left [ \,\left [ Y_{J+s} (\hat {\bf r}) \otimes 
\sigma \right ]_J \otimes Y_{J_i}(\hat {\bf r})\right ]_{J_fM_f} \nonumber
\end{eqnarray}
\noindent where ${\bf r}$ is now the nucleon spatial coordinate and $\hat J =
\sqrt{2J +1}$.

By using Racah algebra to compute the reduced matrix elements of the
operators in Eqs.~(\ref{eq:red-u-mag}-\ref{eq:red-u-el}) we
get finally:
\begin{eqnarray}
\langle ph^{-1};J_f\|\,T^{mag}_J(q)\,\|J_i\rangle
&=&i\frac{f}{m_{\pi}}\left( 1 + \frac{m_\pi}{2M} \right) 
\frac{1}{\sqrt{E^\pi_{n J_i}}}
   \frac{1}{4\pi}\delta_{t_p,-\frac12}\delta_{t_h,\frac12}
   (-1)^{j_p-1/2+J_f}
   \nonumber\\
& &\times \hat{j_p} \hat{j_h} \hat{J} \hat{J_i} \hat{J_f}
          \xi(l_p+l_h+J+J_i)
          \tresj{j_p}{j_h}{J_f}{\frac12}{-\frac12}{0}
   \nonumber\\
&& \times
    \frac{\kappa_p+(-1)^{l_p+l_h+J_f+1}\kappa_h}{\sqrt{J_f(J_f+1)}}
           \tresj{J_f}{J}{J_i}{1}{-1}{0}\nonumber\\
&& \times          \int_0^{\infty}dr\,r^2 R_p(r)j_J(qr)R_{\pi}(r)R_h(r) \\[1cm]
\langle ph^{-1};J_f\|\,T^{el}_J(q)\,\|J_i\rangle
&=&\frac{f}{m_{\pi}}\left( 1 + \frac{m_\pi}{2M} \right) 
\frac{1}{\sqrt{E^\pi_{n J_i}}}
   \frac{1}{4\pi}\delta_{t_p,-\frac12}\delta_{t_h,\frac12}
   (-1)^{j_p-1/2+J_f}
    \nonumber\\
& &\times \hat{j_p} \hat{j_h} \hat{J} \hat{J_i} \hat{J_f}
          \xi(l_p+l_h+J+J_i+1)
          \tresj{j_p}{j_h}{J_f}{\frac12}{-\frac12}{0}
   \nonumber\\
&&\times 
  \left\{
   \frac{\kappa_p+(-1)^{l_p+l_h+J_f+1}\kappa_h}{\hat{J}^2\sqrt{J_f(J_f+1)}}
           \tresj{J}{J_f}{J_i}{1}{-1}{0}
  \right.
  \nonumber\\
&&\times \int_0^{\infty}dr\,r^2 R_p(r)[Jj_{J+1}(qr)-(J+1)j_{J-1}(qr)]
R_{\pi}(r)R_h(r) 
   \nonumber\\
&&  - \sqrt{J(J+1)}
           \tresj{J}{J_f}{J_i}{0}{0}{0}\nonumber\\
&& \times     \left.\frac{1}{q}
     \int_0^{\infty}dr\,r R_p(r)j_{J}(qr)R_{\pi}(r)R_h(r) 
    \right\}.
\end{eqnarray}
where we have defined $\kappa_\alpha=(l_\alpha-j_\alpha)(2j_\alpha+1)$
and the parity function $\xi(n)=1 (0)$ for even (odd) values of $n$.

\end{document}